\documentclass[%
 aps,
 prd,
 floatfix,
 amsmath,amssymb,
 showpacs,
 showkeys,
 reprint,%
]{revtex4-1}

\usepackage{graphicx}
\usepackage{dcolumn}
\usepackage{bm}

\begin{document}
\title{Magnetic properties of the nucleon in a uniform background field}

\author{Thomas Primer}
\affiliation{Special Research Centre for the Subatomic Structure of Matter, School of Chemistry and Physics, University of Adelaide, SA 5005, Australia}
\email{thomas.primer@adelaide.edu.au}

\author{Waseem Kamleh}
\affiliation{Special Research Centre for the Subatomic Structure of Matter, School of Chemistry and Physics, University of Adelaide, SA 5005, Australia}

\author{Derek Leinweber}
\affiliation{Special Research Centre for the Subatomic Structure of Matter, School of Chemistry and Physics, University of Adelaide, SA 5005, Australia}

\author{Matthias Burkardt}
\affiliation{Department of Physics, New Mexico State University, Las Cruces, NM 88003-8001, USA}

\date{\today}

\begin{abstract}
We present results for the magnetic moment and magnetic polarisability of the neutron and the magnetic moment of the proton.
These results are calculated using the uniform background field method on $32^3\times 64$ dynamical QCD lattices provided by the PACS-CS collaboration as part of the ILDG.
We use a uniform background magnetic field quantised by the periodic spatial volume.
We investigate ways to improve the effective energy plots used to calculate magnetic polarisabilities, including the use of correlation matrix techniques with various source smearings.
\end{abstract}


\maketitle

\section{Introduction}\label{intro}

The magnetic moment and magnetic polarisability are fundamental properties of a particle that describe its response to an external magnetic field.
Developing the ability to calculate these properties via the first principles approach of lattice QCD is important.
There are two well known techniques for calculating magnetic moments on the lattice.
One is the three-point function method \cite{Martinelli:1988,Draper:1989,Leinweber:1990}, which is used to calculate baryon electromagnetic form factors that can be converted into magnetic moments by performing an extrapolation to zero momentum.
The other is the background field method \citep{Bernard:1982,Smit:1987,Rubinstein:1995,Burkardt:1996,Zhou:2003,Lee:2005,Lee:2006,Lee:2008}, which uses a phase factor on the gauge links to induce an external field across the whole lattice.
This external field causes an energy shift from which the magnetic moment and polarisability can be derived by making use of  the following energy-field relation \cite{Bernard:1982,Fiebig:1988},
	\begin{equation}\label{erelation}
	E(B) = M_N - \vec\mu \cdot \vec{B} + \frac{e\vert{B}\vert}{2M_N} - \frac{4\pi}{2}\beta B^2 + \mathcal{O}(B^3),
	\end{equation}
defining $\vec{\mu}$ as the magnetic moment and $\beta$ as the magnetic polarisability.
We note the term $e\lvert{B}\rvert/2M_N$ is the ground state Landau energy.
In principle, there is a tower of energy levels with energy, $(2n+1)e\lvert{B}\rvert/2M_N$ for $n=0,1,2,...$ .

When deriving the background field method on a periodic lattice there arises a quantisation condition which limits the available choices of magnetic field strength based on the size of the lattice \cite{Smit:1987}.
If the lattice is too small the field will be large and higher order terms in the energy relation of Eq.~\eqref{erelation} will begin to dominate \cite{Burkardt:1996}.
Previous calculations have avoided this problem by using a Dirichlet boundary condition in a spatial dimension and a linearised form of the phase factor, which allows for an arbitrary choice of field strength \cite{Lee:2008}.
Others have used the exponential phase, but instead of correcting the value of the field at the boundary they put the quark origin at the centre of the lattice and hope that the boundary is far enough away for the effects of the discontinuity to be small \cite{Rubinstein:1995}.
Using either of these methods introduces finite volume errors which can be hard to predict.
Our calculation is the first to use periodic boundary conditions and the quantised exponential phase factor, creating a uniform magnetic field everywhere.
We present results for both the magnetic moment and the magnetic polarisability of the neutron.
For the proton we present only magnetic moment results because the Landau levels interfere with polarisability calculations for charged particles.

\section{Background Field Method}\label{method}

We make use of the background field method to simulate a constant magnetic field along one axis \cite{Smit:1987}.
The technique is formulated on the lattice by first considering the continuum case, where the covariant derivative is modified by the addition of a minimal electromagnetic coupling,
\begin{equation}
D_{\mu} = \partial_{\mu}+gG_{\mu}+qA_{\mu},
\end{equation}
where $A_{\mu}$ is the electromagnetic four-potential and $q$ is the charge on the fermion field.
On the lattice this is equivalent to multiplying the usual gauge links by a simple phase factor
\begin{equation}
U_{\mu}^{(B)}(x)=\exp(iaqA_{\mu}(x)).
\end{equation}

To obtain a uniform magnetic field along the z-axis we note that $\vec{B} = \vec{\nabla} \times \vec{A},$ and hence 
\begin{equation}\label{Bz}
B_z = \partial_x A_y - \partial_y A_x.
\end{equation}
Note that this equation does not specify the gauge potential uniquely, there are multiple valid choices of $A_\mu$ that give rise to the same field.
We choose $A_x = -By$ to produce a constant magnetic field of magnitude $B$ in the $z$ direction.
The resulting field can be checked by examining a single plaquette in the $(\mu,\nu) = (x,y)$ plane, which is related to the magnetic field through the field strength tensor,
\begin{equation}
\Box_{\mu\nu}(x) = \exp\left(iqa^2F_{\mu\nu}(x)\right),
\end{equation}
which is exact for a constant background field because all higher order terms involve a second or higher order derivative.
For a general plaquette at coordinates $x,y$ the result is,
\begin{equation}
\exp(-iaqBy)\exp(iaqB(y+a)) = \exp(ia^2qB),
\end{equation}
giving the desired field over most of the lattice.
However on a finite lattice $(0 \le x/a \le N_x-1), (0 \le y/a \le N_y-1)$ there is a discontinuity at the boundary due to the periodic boundary
conditions.
In order to fix this problem we make use of the $\partial_x A_y$ term from equation \eqref{Bz}, giving $A_y$ the following values,
\begin{equation}
A_y(x,y) = \begin{cases} 0, & \mathrm{for\ } y/a < N_y-1, \\
	           N_y Bx, & \mathrm{for\ } y/a = N_y-1. \end{cases}.
\end{equation}
This ensures that we now get the required value at the $y/a=N_y-1$ boundary.

There is then the issue of the double boundary, $x/a = N_x-1$ and $y/a=N_y-1$, where the plaquette only has the required value under the condition
$\exp(-ia^2 qBN_x N_y)=1$.
This gives rise to the quantisation condition which limits the choices of magnetic field strength based on the lattice size,
\begin{equation} \label{quantisationcondition}
qBa^2 = \frac{2\pi n}{N_x N_y},
\end{equation}
where $n$ is an integer specifying the field strength in multiples of the minimum field strength quantum.

\section{Simulation Details}\label{details}

These calculations use the 2+1 flavour dynamical-fermion configurations provided by the PACS-CS group  \citep{PACS-CS} through the ILDG \citep{ILDG}.
These are $32^3 \times 64$ lattices using a clover fermion action and Iwasaki gauge action with $\beta=1.9$ and physical lattice spacing $a=0.0907(13)$ fm.
We use four values of the light quark hopping parameter, $\kappa_{ud} =$ 0.13700, 0.13727, 0.13754, 0.13770, corresponding to the pion masses $m_\pi $= 702, 572, 413, 293 MeV.
The lattice spacing for each mass was set using the Sommer scale with $r_0=0.49$ fm.
The size of the ensemble was 320 for the two lighter masses and 400 for the heavier ones.

In order to get correlation functions at four different magnetic field strengths we calculated propagators at six non-zero field strengths, $qBa^2 = $+0.0061, $-$0.0123, +0.0184, +0.0245, $-$0.0368, $-$0.0492.
These correspond to $n=+1,-2,+3,+4,-6$ and $-8$ in Eq.~\eqref{quantisationcondition}.
Using the relationships $q_d = -e/3$ and $q_u = 2e/3$ to combine up and down quark propagators with the appropriate field strengths resulted in hadrons in fields of strength $eB = $ $-$0.087, +0.174, $-$0.261, $-$0.345 GeV$^2$ at the physical lattice spacing.
Unless specified otherwise we used the interpolating field $\chi_1 = (u^TC\gamma_5d)u$ with 100 sweeps of Gaussian smearing at the source.
We put the origin of the electromagnetic gauge field at the same lattice site as the quark origin to ensure that the smeared source maintains good overlap with the ground states.

It should be noted that the configurations are dynamical only in the QCD sense, there was no magnetic field included when they were generated.
The background field can be put on the sea quarks by performing a separate HMC calculation for each field strength, but this is obviously very computationally expensive.
It also destroys the correlations between the different field strengths which would lead to much larger errors in the energy shifts used to calculate moments and polarisabilities.
While techniques for a re-weighting of configurations in order to correct for the background field are under exploration \cite{Freeman:2012}, these have not been employed in this work.
Because these effects are proportional to $SU(3)$ flavour symmetry breaking in the vacuum we anticipate that the corrections due to the effect of the background field on the sea quarks will be small.

We also performed an initial calculation using quenched gauge configurations.
These were $32^3\times 40$ lattices using a FLIC fermion action and Symanzik improved gauge action at $\beta=4.52$.
There were 192 configurations at seven quark masses, corresponding to $m_\pi=$ 0.8400, 0.7745, 0.6929, 0.6261, 0.5399, 0.4353, 0.2751 GeV.
The lattice spacing was $a=0.128$ fm and like the dynamical calculation, boundary conditions were periodic for the spatial dimensions and fixed for the time boundary.
We used fields corresponding to $n=1,\,-2,\,4,\,-8$ in the quantisation condition to save on computation.

\section{Magnetic Moment}\label{moment}

\subsection{Formalism}

When a charge or system of charges with angular momentum is placed in an external magnetic field it is energetically favourable to have its axis either aligned or anti-aligned with the direction of the field.
The tendency of the system to align with the field is proportional to the magnetic moment of the system and the strength of the field.

We calculate zero-momentum projected correlation functions containing spin-up and spin-down components in the (1,1) and (2,2) positions of the Dirac matrix respectively.
For a magnetic field aligned to the axis of the spin we see the magnetic moment manifest as a shift in the energy which has the same magnitude, but opposite sign, for spin-up and spin-down.

We make use of the sign difference in the energy shift between spin-up and spin-down in order to isolate the magnetic moment term from the expansion of the energy.
Taking the difference of the spins,
\begin{equation}\label{difference}
\delta E(B) = \frac{1}{2}\left(E_\uparrow(B) - E_\downarrow(B)\right) = -\mu B.
\end{equation}
In addition to the bare mass and polarisability term, this difference also cancels out the Landau energy term $e\lvert{B}\rvert/2M_N$.
For the neutron this term should be zero because it is proportional to the net charge.
However for the proton, even though taking the difference cancels out the term, it can still affect the results.
This is because we use a standard projection to zero momentum in our correlation functions, but when Landau levels are present one obtains a superposition of Landau states.
There are proposed techniques for dealing with the Landau levels \cite{Tiburzi:2012}, but we have found that the effect on the magnetic moment results is small, and defer this issue to a subsequent investigation.

In terms of correlation functions there are multiple valid ways of taking the spin-difference, for example fitting the energy and then taking the difference or combining correlation functions and then fitting.
By taking a combination of correlation functions before fitting for the energy the statistical error is greatly reduced and provides strong constraints on the fit regime.
This is because the errors are highly correlated between the zero and non-zero field correlation functions, meaning the fluctuations do not change significantly due to the field.
The combination required for isolating the moment term can be written as,
\begin{equation}
\delta E(B,t) = \frac{1}{2}\left(\ln\left(\frac{G_{\uparrow}(B,t)}{G_{\uparrow}(0,t)}\frac{G_{\downarrow}(0,t)}{G_{\downarrow}(B,t)}\right)\right)_{\rm fit}.\label{spindiffeq}
\end{equation}
The inclusion of the bare correlation functions without a magnetic field in this expression is not strictly necessary, but it is useful in correcting for the small statistical difference between spin-up and spin-down zero-field energies and making the zero field point zero by construction.
We also define spin-up to mean aligned with the magnetic field and spin-down to mean anti-aligned to the field so that we can treat all the fields as positive in our discussion.

\subsection{Results}

\begin{figure}\centering
\includegraphics[width=0.30\textwidth,angle=90]{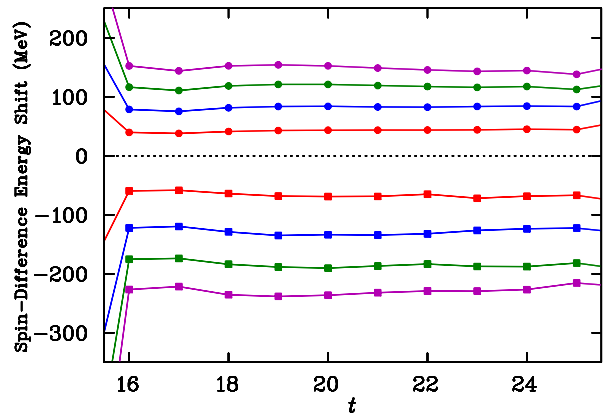}
\caption{Spin-difference energy shift of Eq.~\eqref{spindiffeq} for the heaviest quark mass at all four field strengths. The proton values are given by the squares and the neutron values by the circles. The shifts increase in magnitude with the strength of the field.}
\label{diffplots}
\end{figure}

Figure \ref{diffplots} shows the energy shift from the difference of spin-up and spin-down nucleons for the heaviest quark mass at all four non-zero magnetic field strengths.
Both the proton and the neutron show a good linear progression over the field strengths as expected, with excellent plateaus.
There is very little excited state contribution to the energy shifts in evidence.
The neutron effective energy is generally slightly smoother than the proton, with similar results for the other quark masses.
The larger errors in the proton energies may be due to the effect of the Landau levels in the momentum projection.
At the two higher field strengths there tends to be a small drift in the value over time, with the true plateau perhaps only occurring at around time slice 23 or 24.
This leads to a slight difference in the value of the energy shift depending on the choice of fit window, however this has little effect on the magnetic moment result for reasons described below.

\begin{figure}\centering
\includegraphics[width=0.3\textwidth,angle=90]{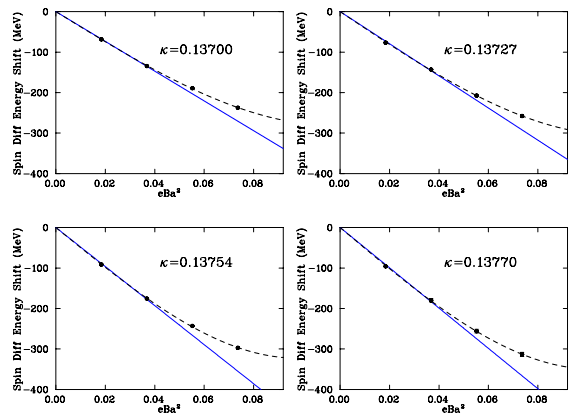}
\caption{Fits of the spin-difference energy shift to the field strength at each quark mass for the proton. The solid line is a purely linear fit to just the first two points and the dashed line is a linear plus cubic fit to all four points.}
\label{difffits}
\end{figure}

\begin{figure}\centering
\includegraphics[width=0.3\textwidth,angle=90]{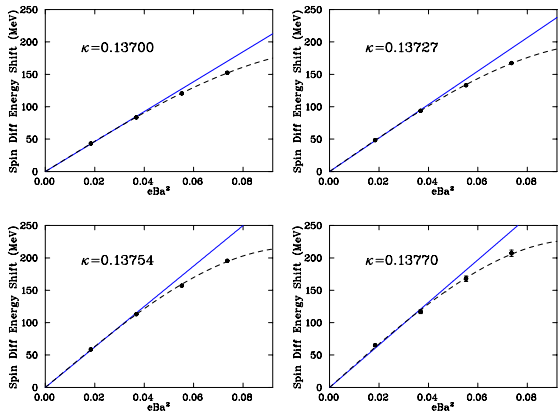}
\caption{Fits of the spin-difference energy shift to the field strength at each quark mass for the neutron. The solid line is a purely linear fit to just the first two points and the dashed line is a linear plus cubic fit to all four points.}
\label{difffits2}
\end{figure}

Figures \ref{difffits} and \ref{difffits2} show the spin-difference energy shifts plotted against the magnetic field strength.
These are fit to a linear coefficient which gives the magnetic moment.
In order to fit the largest field strength, and to a lesser extent the second largest, we had to include a cubic term in the fit.
With the cubic term included all four data points are fit easily.
The cubic term is able to absorb some variation in the energy shifts at the higher field strengths, which is why the drift in the effective energy shown in Fig.~\ref{diffplots} doesn't significantly affect the resulting magnetic moment value.

This is seen in Table~\ref{windowtable}, which gives values of the neutron magnetic moment for a number of fit windows.
The same window is used at every field strength in order to maintain consistency and prevent introducing systematic errors.
The values agree well within errors for all but the earliest fit window, suggesting that time slice 19 is slightly too early to fit due to excited state contamination.
This shows that the first two points are the main drivers of the linear coefficient and therefore the magnetic moment value.
We also performed a purely linear fit to only the first two points and found that the linear coefficients agreed well within errors.
Since the fit is naturally constrained to go through zero, two non-zero field points are enough to give us confidence that our field strengths are small enough to make the higher order contributions negligible.

\begin{figure}\centering
\includegraphics[width=0.35\textwidth,angle=90]{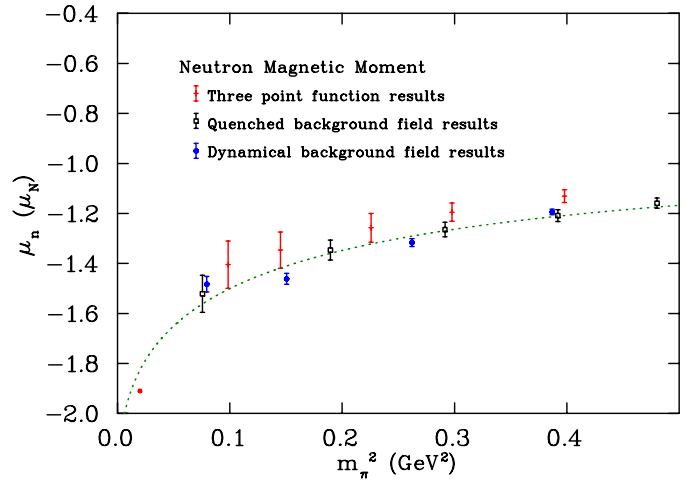}
\caption{Neutron magnetic moment as a function of pion mass squared. The left most point gives the experimental value \cite{PDG:2012}. The dashed line is a chiral extrapolation of the dynamical points.}
\label{Nmomentplot}
\end{figure}
\begin{figure}\centering
\includegraphics[width=0.35\textwidth,angle=90]{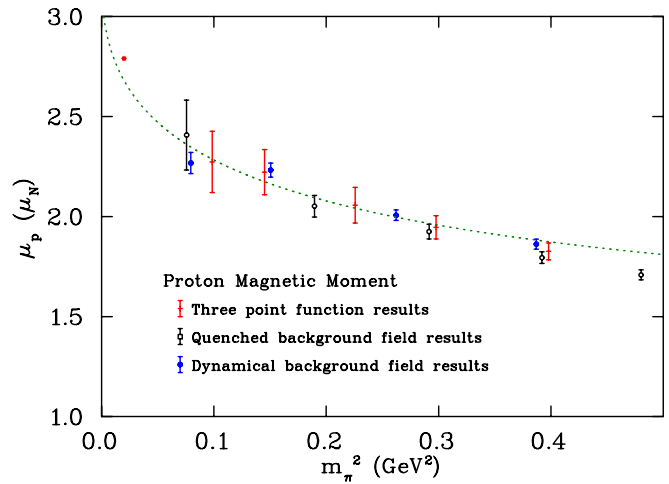}
\caption{Proton magnetic moment as a function of pion mass squared. The left most point gives the experimental value \cite{PDG:2012}. The dashed line is a chiral extrapolation of the dynamical points.}
\label{Pmomentplot}
\end{figure}


Figures \ref{Nmomentplot} and \ref{Pmomentplot} show the proton and neutron magnetic moment results, compared with a three point function calculation for reference \citep{Boinepalli:2006}.
Here we used values taken from fit window 20-22 because this window had good $\chi^2$ per degree of freedom and small errors.
The magnetic moment is reported in units of nuclear magnetons, which are reached by,
\begin{equation}
\mu = -\frac{\delta E}{eB}\left[\frac{e}{2M_N}\right]2M_N
\end{equation}
where we have started with Eq.~\eqref{difference} and introduced the elementary charge $e$ since we actually fit the energy shift against $eB$, then bring in twice the physical nucleon mass $M_N$ in order to get the nuclear magneton ($\mu_N = \frac{e\hbar}{2M_N}$), given that we are using natural units ($c=\hbar=1$).

The results compare favourably.
The lines are chiral fits to the dynamical results using the approach from \citep{Leinweber:1999}, and guide the anticipated trajectory to the physical point.
The reason the extrapolated values are smaller in magnitude than the experimental values is expected to come from finite volume effects \cite{Hall:2012} as those have not been examined here.


\begin{table}[b!]\vspace{-24pt}
\begin{ruledtabular}
\caption{Magnetic moment values for the neutron at each $\kappa$ value for a variety of fit windows.}\label{windowtable}
\begin{tabular}{c | c c c c}
 window & 0.13700 & 0.13727 & 0.13754 & 0.13770 \\ \hline
19-21  & -1.187(12) & -1.300(13) & -1.420(16) & -1.486(36) \\ 
20-22  & -1.194(11) & -1.317(15) & -1.462(22) & -1.483(30) \\ 
21-23  & -1.198(13) & -1.338(20) & -1.454(27) & -1.500(40) \\ 
22-24  & -1.201(15) & -1.343(25) & -1.454(32) & -1.508(49) \\ 
20-24  & -1.199(10) & -1.321(15) & -1.462(20) & -1.485(31) \\ 
\end{tabular}
\end{ruledtabular}
\end{table}

\section{Magnetic Polarisability}\label{polarisability}

\subsection{Formalism}

The magnetic polarisability is a measure of the deformation of a non-pointlike particle when it is placed in a magnetic field.
This deformation causes a change in the energy which we can measure using the background field method.
The effect of the magnetic polarisability is second order in $B$.
This means that at the ``small'' field strengths we are using the effect is much smaller than that due to the magnetic moment, which can make it hard to measure.
It also makes it more important to use the full exponential phase factor, since the errors introduced by the linearised form are also at order $B^2$ \cite{Guerrero:2008}.


To extract the polarisability from the energy we take the average of spin-up and spin-down energy shifts to remove the magnetic moment term and explicitly subtract the zero-field mass.
The spin-averaged energy shift is
\begin{eqnarray*}
\delta E_\beta(B) &=& \frac{1}{2}\big((E_\uparrow(B)-E_\uparrow(0)) + (E_\downarrow(B)-E_\downarrow(0))\big) \\
				 &=& \frac{e\vert{B}\vert}{2M_N}- \frac{4\pi}{2}\beta B^2.
\end{eqnarray*}
This leaves us with the polarisability term, but also with another term due to the Landau energy.
This energy arises from the quantisation of orbits for charged particles in magnetic fields and can't be isolated from the relevant polarisability term.
As a result it is difficult to calculate magnetic polarisabilities of charged particles because there is not only the ground state Landau energy but also a tower of Landau levels with energy $(2n+1)\frac{e\lvert{B}\rvert}{2M_N}$.
The need for small field strengths makes the Landau level problem even worse because it means the Landau levels are closer together, which makes it take longer in Euclidean time for the levels above the ground state to be exponentially suppressed \citep{Tiburzi:2012}.

\begin{figure}\centering
\includegraphics[width=0.30\textwidth,angle=90]{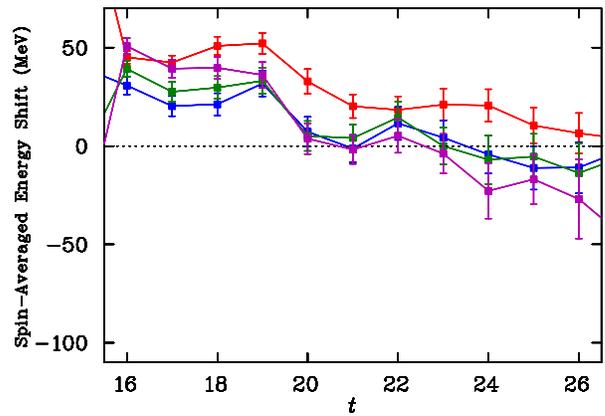}
\caption{Proton spin-averaged energy shift for the heaviest quark mass at all field strengths. The top line is the smallest field strength, with the other three agreeing well within errors for most of the relevant time frame.}
\label{Landau}
\end{figure}

The influence of the Landau levels on the proton is readily apparent in Figure~\ref{Landau}, which shows the spin-average of the energy shift due to the field.
Since the experimental value of the magnetic polarisability is approximately the same for the proton and neutron we would expect this to look similar to the neutron results in Fig.~\ref{sumplots}.
Instead we see much larger errors and no consistent trend across the field strengths.
Due to the large and unpredictable systematic errors caused by this effect we are not presenting values for the magnetic polarisability of the proton in this first exploratory investigation.
Fortunately for a neutral particle like the neutron the Landau term is zero and can be ignored.

As with the magnetic moment we construct ratios of correlation functions which we then fit for an effective energy,
\begin{equation}\label{polratio}
\delta E_\beta(B,t) = \frac{1}{2}\left(\ln\left(\frac{G_{\uparrow}(B,t)}{G_{\uparrow}(0,t)}\frac{G_{\downarrow}(B,t)}{G_{\downarrow}(0,t)}\right)\right)_{\rm fit}.
\end{equation}
In this case the zero-field correlators are necessary to remove the bare neutron mass.
Combining the correlation functions before fitting is especially important in the polarisability case because the energy shift is smaller than the errors on the zero-field mass.
This means that if the correlated errors were not allowed to cancel before the fit we would not see a clear signal for the shift due to the polarisability.

\subsection{Results}

\begin{figure}\centering
\includegraphics[width=0.3\textwidth,angle=90]{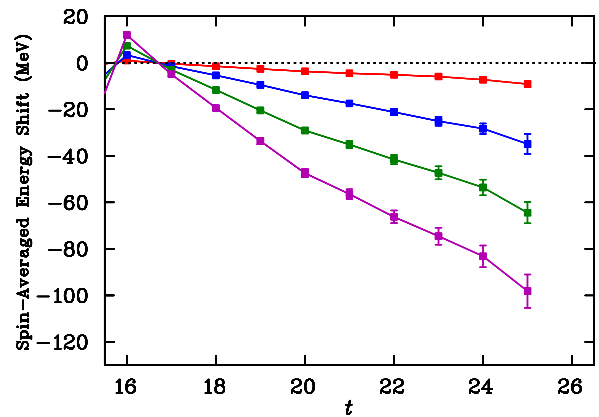}
\caption{Neutron spin-averaged energy shift for the heaviest quark mass at all four field strengths. The magnitude of the shift increases with the field strength.}
\label{sumplots}
\end{figure}

\begin{figure}\centering
\includegraphics[width=0.3\textwidth,angle=90]{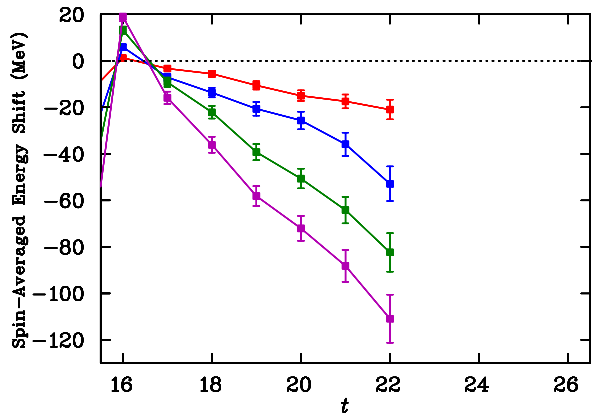}
\caption{Neutron spin-averaged energy shift for the lightest quark mass at all four field strengths. The magnitude of the shift increases with the field strength.}
\label{Sum13770}
\end{figure}

\begin{figure}\centering
\includegraphics[width=0.3\textwidth,angle=90]{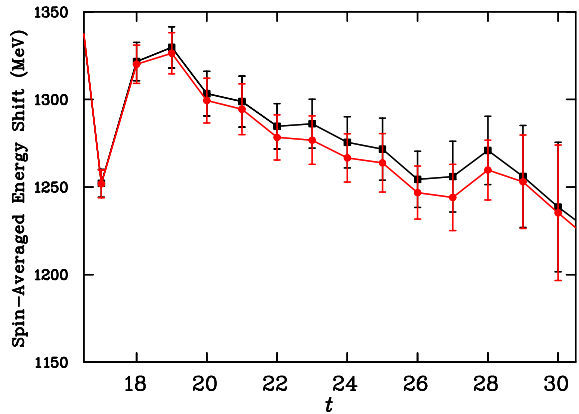}
\caption{Spin-averaged effective mass for the neutron at $\kappa=$ 0.13727. The top line (squares) is for zero magnetic field and the bottom line (circles) is with the smallest field strength considered.}
\label{logG}
\end{figure}

Figure \ref{sumplots} gives the spin-averaged effective energy shift for the heaviest quark mass considered.
Unlike the spin-difference case the plateau behaviour is quite poor, with a fairly constant downward slope that begins to plateau only after significant noise is appearing.
Only in the case of the smallest field strength does something like a real plateau appear before the signal is lost to noise.
The situation is very similar at other quark masses, with the errors getting larger and the noise coming earlier at lighter quark masses, as seen in Figure~\ref{Sum13770} for the lightest quark mass considered.
The plots show that at each field strength the energy shift starts at approximately zero and grows with Euclidean time.
Typically the lack of a plateau in an effective energy plot is due to the presence of excited state energies in addition to the ground state.
Figure~\ref{logG}, illustrating the bare effective mass, does reveal a systematic drift in the energy suggesting some improvement in the interpolating field may be possible.

\begin{figure}[b!]\centering
\includegraphics[width=0.3\textwidth,angle=90]{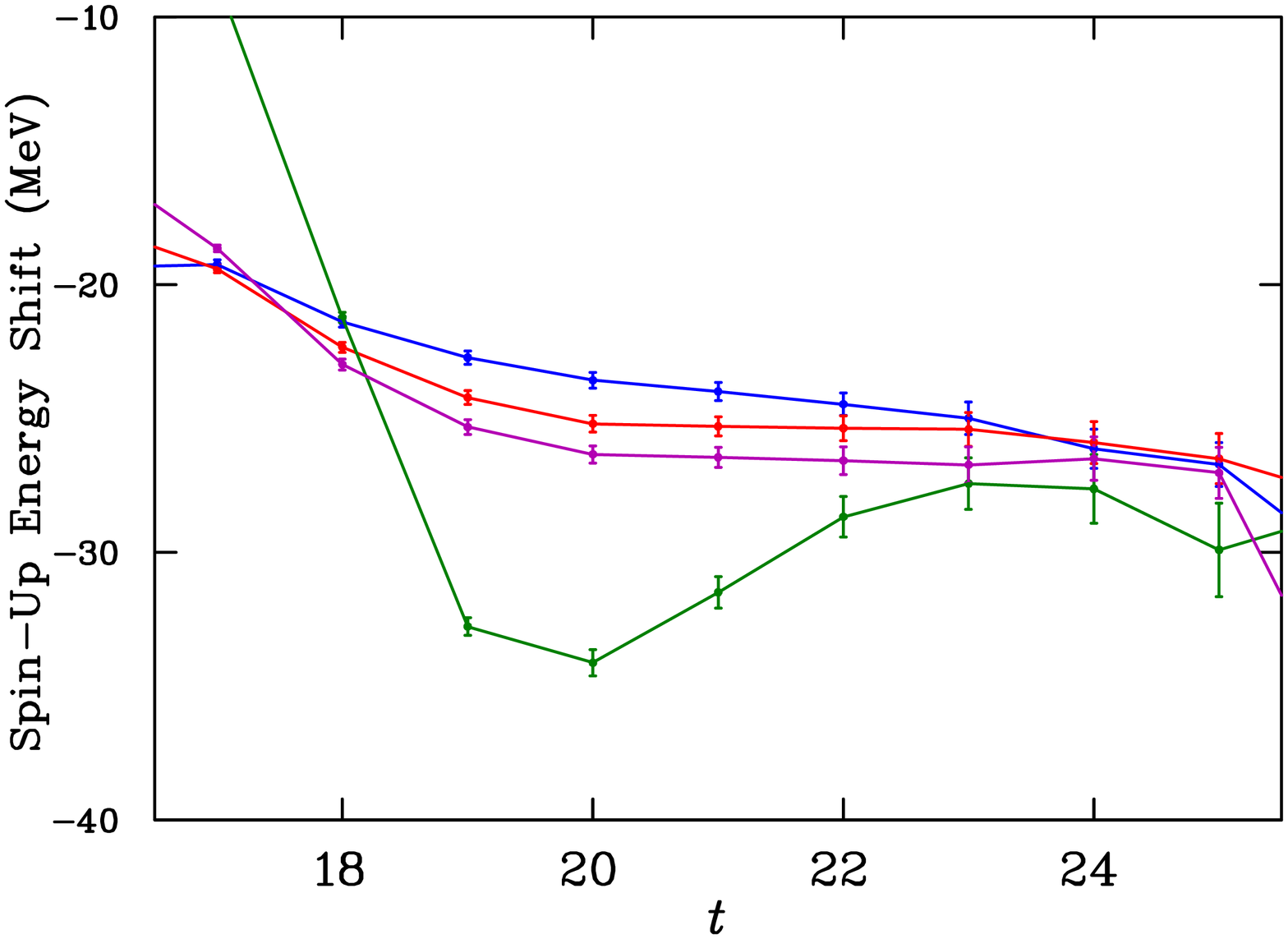}
\includegraphics[width=0.3\textwidth,angle=90]{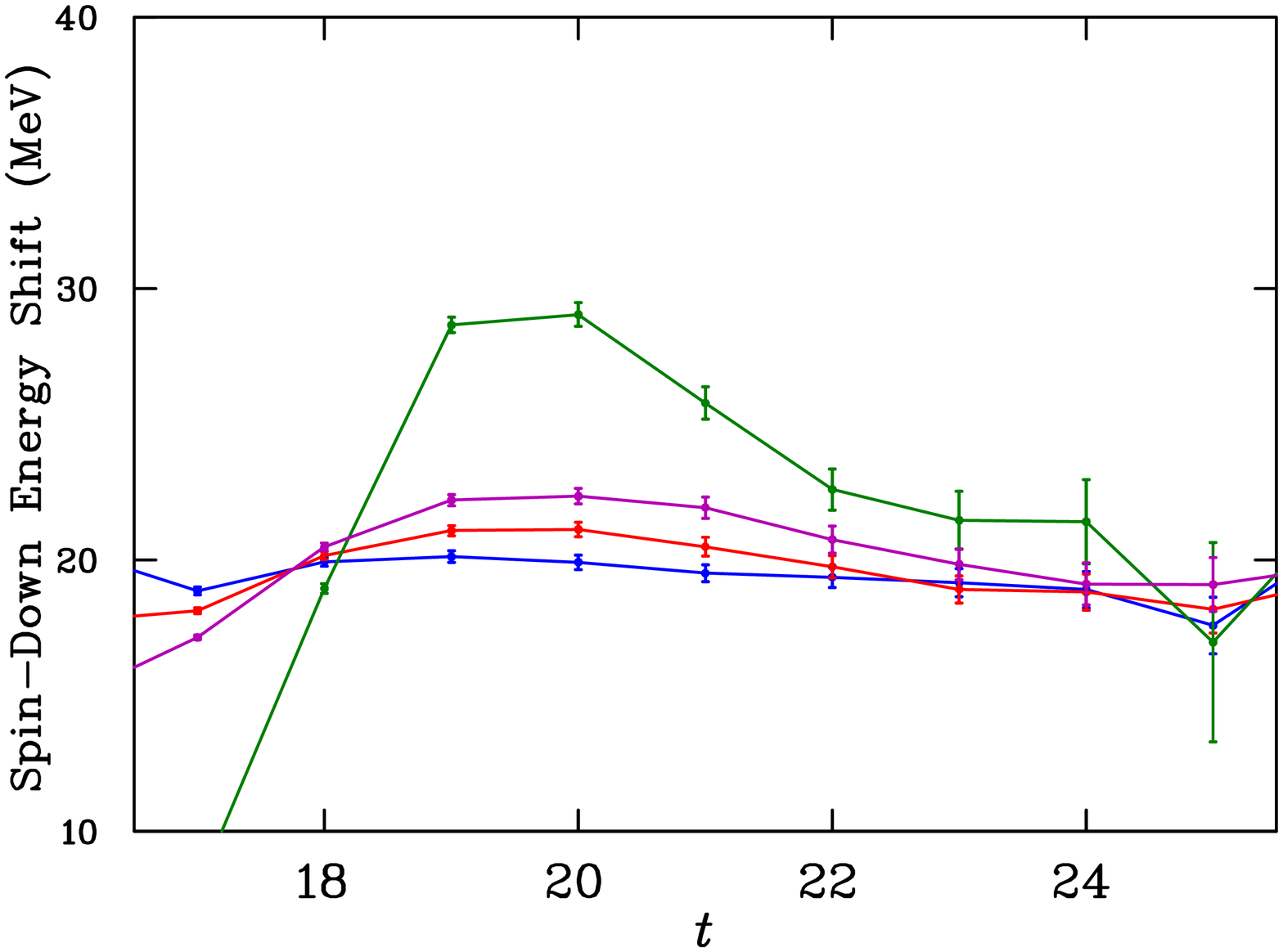}
\caption{Energy shift at the smallest field strength. For spin-up we have from top to bottom: 100, 35, 16 sweeps of smearing and point source, for spin-down the order is reversed.}
\label{upcompare}
\end{figure}

In order to check for excited state overlap and to try and improve the plateau behaviour we looked at different sources.
We experimented with a number of different source smearings, trying 16 and 35 sweeps in addition to our usual 100.
We also tried a point source on the basis that it should have no bias towards any shape and may therefore reach the required form more quickly.
Figure \ref{upcompare} shows the energy shift due to the field for all smearing choices at the heaviest quark mass and the smallest field.
The three smeared sources have different behaviour but agree well within errors by time slice 24, just before the signal is lost to noise.
The point source has large excited state contributions and approaches agreement with the other sources as the signal is lost.

\begin{figure}\centering
\includegraphics[width=0.3\textwidth,angle=90]{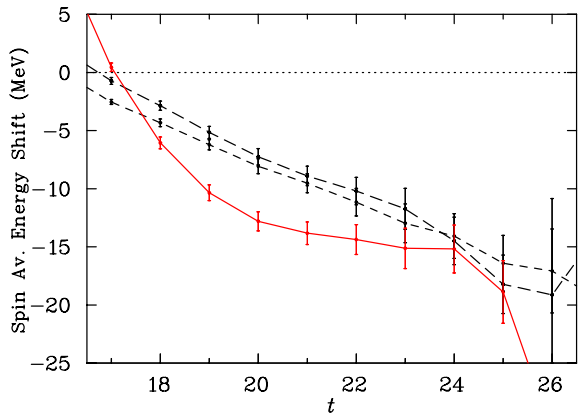}
\caption{Spin-averaged effective energy shift for the heaviest quark mass at the smallest background field. The dashed lines are for 16 and 100 sweeps of smearing and the solid line is from the combination of spin-up with 16 sweeps and spin-down with 100 sweeps.}
\label{sumcompare}
\end{figure}

We notice from these plots that the best plateau behaviour comes from 16 sweeps of smearing for spin-up and 100 sweeps of smearing for spin-down.
To take advantage of this we constructed a spin-average from the spin-up correlation function with 16 sweeps and the spin-down correlation function with 100 sweeps.
We found the improved plateau behaviour shown in Figure \ref{sumcompare}, leading us to investigate further the possibility of combining different source smearings.

The variational method as implemented in Ref.~\cite{Mahbub:2011} involves using an $n \times n$ correlation matrix $G_{ij}(t)$ constructed from different source and sink smearing levels to solve a pair of eigenvalue equations.
The right and left eigenvectors $u_j^\alpha$ and $v_i^\alpha$ can then by used to project out energy eigenstates $\alpha$ to effectively isolate the $n-1$ lowest energy states,
\begin{equation}\label{eigenvector}
G^\alpha(t) = v_i^\alpha G_{ij}(t)u_j^\alpha
\end{equation}
Combining correlation matrix techniques with the background field method introduces new considerations.
Normally the eigenvector analysis is performed on spin-averaged correlation functions because spin-up and spin-down are equal up to statistics.
For a background field calculation we need to consider spin-up and spin-down separately, with each field strength getting its own eigenvector equation to solve.
This is because the Hilbert space changes and one must first isolate the state before combining it with states from other Hilbert spaces.
In other words, the eigenvectors $u_i^\alpha$ and $v_j^\beta$ are field and spin dependent and a recurrence relation leading to a generalised eigenvalue equation cannot be written for combinations of spins and fields.
After solving the eigenvalue equations we construct the same ratio as in Eq.~\eqref{polratio} but using the projected correlation functions from Eq.~\eqref{eigenvector}.

We first performed a variational analysis using a $2\times 2$ correlation matrix made from 16 and 100 sweeps of smearing.
The resultant spin-average energy shift using the ground state projected correlation functions was approximately equal to the original 16 and 100 sweeps correlation functions.
The plateau behaviour was not noticeably better than using either of the smearings alone.
We also tried other various combinations of sources with different smearings including 16, 35 and 100 sweeps as well as different interpolating fields in $3\times 3$ and $4\times 4$ correlation matrices.
None of these combinations was found to result in a statistically significant improvement in the plateau behaviour.

One hypothesis is that the neutron is not as free from Landau level effects as we believed.
The idea is that there is enough play in the extended structure of the neutron to allow the $u$ and $d$-quarks to respond to the external field with a non-trivial Landau energy.
Since the neutron has a non-zero charge radius it may have a Landau energy but with a small effective charge.
This would lead to very closely spaced Landau levels which could decay smoothly in the manner we observe.


Since we could not achieve good plateaus for fitting, we extracted energy shifts by simply taking the value at the point just before noise dominates the signal.
Because all the different smearings agree at this point we are confident the excited states have been suppressed.

\begin{figure}\centering
\includegraphics[width=0.3\textwidth,angle=90]{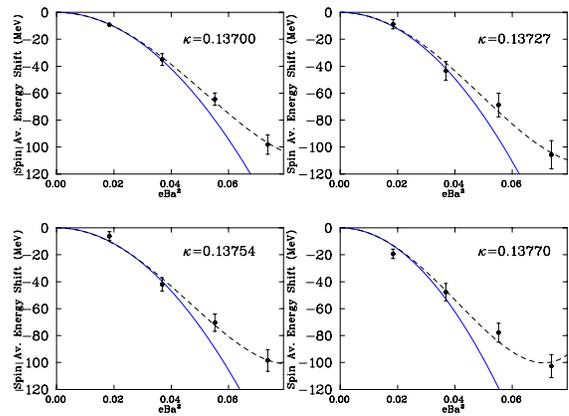}\\
\caption{The spin-averaged energy shifts as a function of the background field strength for the neutron. The solid line is a pure quadratic fit to the first two points while the dashed line is a fit of all the points to a quadratic plus quartic.}
\label{sumfits}
\end{figure}

Figure \ref{sumfits} shows fits of these spin-averaged energy shifts as a function of the field strength.
In addition to the quadratic polarisability term we required a quartic term in order to fit the higher field strengths.
This higher order term is small at the heaviest mass but starts to become significant at the lighter masses.

\begin{figure}\centering
\includegraphics[width=0.35\textwidth,angle=90]{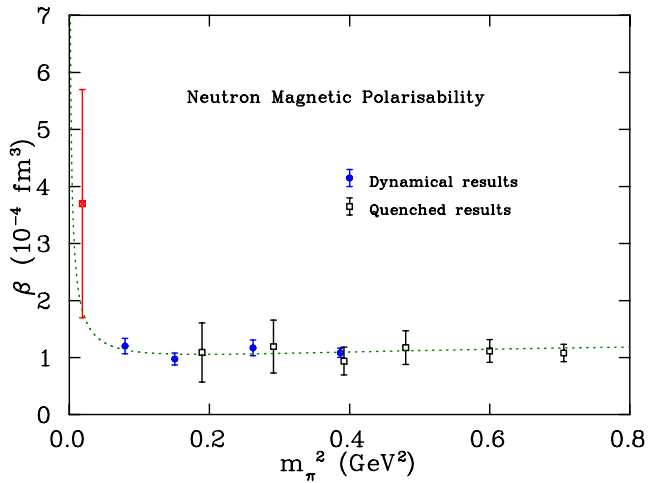}
\caption{Neutron magnetic polarisability vs pion mass. The red point illustrates the experimental value \cite{PDG:2012}. The line represents a fit of the dynamical points using $\chi$PT.}
\label{polarisabilityplot}
\end{figure}

Figure \ref{polarisabilityplot} shows our neutron magnetic polarisability results with a comparison between our quenched and dynamical calculations.
The quenched and dynamical results agree well within errors.
The dashed line shows a fit of the dynamical results to
\[ \beta = a + b/m_\pi + c\ln(m_\pi) + d m_\pi^2, \]
where the values of coefficients $b$ and $c$ were set from values calculated in $\chi\mathrm{PT}$ \cite{Bernard:1993} and $a$ and $d$ were fit freely.
The extrapolated value of $1.8\pm 0.2 \times 10^{-4}\,\,\mathrm{fm}^3$ is well within the error bar of the experimental value, which is large due to how difficult the measurement is to perform.
With an improved calculation at near physical pion mass we could soon expect a lattice result that verifies the chiral curvature and sets a challenge for the measurement of the experimental value for the neutron magnetic polarisability.

\begin{table}\centering
\begin{ruledtabular}
\caption{Summary of the main results.}\label{resultstable}
\begin{tabular}{c c c c c c}
 & $m_\pi$ & a & $\mu_n$ & $\mu_p$ & $\beta_n$ \\
 $\kappa$ & (MeV) & (fm) & ($\mu_N$) & ($\mu_N$) & ($10^{-4}$ fm$^3$) \\ \hline
0.13700 & 702 & 0.1022(15) & $-$1.19(1) & 1.86(3) & 1.08(8) \\ 
0.13727 & 572 & 0.1009(15) & $-$1.32(2) & 2.01(3) & 1.17(14) \\ 
0.13754 & 413 & 0.0961(13) & $-$1.46(2) & 2.23(4) & 0.98(10) \\ 
0.13770 & 293 & 0.0951(13) & $-$1.48(3) & 2.27(5) & 1.20(13) \\ 
\end{tabular}
\end{ruledtabular}
\end{table}

\section{Conclusion}

We have performed the first calculations of the magnetic moment and magnetic polarisability of the neutron in a uniform background field.
Results for the magnetic moment are very clear and agree well with previous calculations.
The approach can be used in a precision manner to directly determine the magnetic moment without the need for extrapolating form factors in $Q^2$.

Magnetic polarisability calculations have proved more difficult due to late appearing plateaus.
We have calculated results that agree with our previous quenched results and have an excellent approach to the experimental value.
The extrapolated value at the physical pion mass is $1.8\pm 0.2 \times10^{-4}\, \mathrm{fm}^3$.
Here the uncertainty is statistical only.
Future studies in chiral effective field theory and in lattice QCD are needed to quantify and correct for the systematic errors associated with the finite volume of the lattice and problems associated with the isolation of the ground state energy shift.

Further study is required to improve our understanding of the physical effects associated with the magnetic polarisability calculation.
It is important to determine the role of $u$ and $d$-quark sectors within the neutron and elucidate their contributions to the Landau energy of this neutral baryon.

\acknowledgments
We are grateful for the generosity of the PACS-CS collaboration for providing the gauge configurations used in this study.
The contributions of the ILDG in making the configurations accessible is also appreciated.
This research was undertaken with the assistance of resources at the NCI National Facility in Canberra, Australia, and the iVEC facilities at the University of Western Australia.
These resources were provided through the National Computational Merit Allocation Scheme, supported by the Australian Government.
This research is supported by the Australian Research Council.

\bibliography{MomPolPaper}

\end{document}